\def\year{2021}\relax
\documentclass[letterpaper]{article} % DO NOT CHANGE THIS
\usepackage{aaai21}  % DO NOT CHANGE THIS
\usepackage{times}  % DO NOT CHANGE THIS
\usepackage{helvet} % DO NOT CHANGE THIS
\usepackage{courier}  % DO NOT CHANGE THIS
\usepackage[hyphens]{url}  % DO NOT CHANGE THIS
\usepackage{graphicx} % DO NOT CHANGE THIS
\usepackage{natbib}  % DO NOT CHANGE THIS AND DO NOT ADD ANY OPTIONS TO IT
\usepackage{caption} % DO NOT CHANGE THIS AND DO NOT ADD ANY OPTIONS TO IT

\usepackage{amsmath,amssymb,amsfonts}
\usepackage{algorithmic}
\usepackage{graphicx}
\usepackage{textcomp}
\usepackage{xcolor}
\usepackage[utf8]{inputenc}
\usepackage{CJK}
\usepackage{comment}
\usepackage{subfigure}
\usepackage{float}
\usepackage{algorithm}  
\usepackage{amsmath}  
\usepackage{mathrsfs}
\usepackage{amsmath}
\usepackage{amssymb}
\usepackage{makecell}
\usepackage{multirow}

\usepackage{soul}
\usepackage{textcomp} 
\usepackage{bm}
\usepackage{booktabs}
\usepackage{mathrsfs}
\usepackage{amsmath}
\usepackage{etoolbox}
\usepackage[none]{hyphenat}
\makeatletter

\title{Heterogeneous Network Embedding for Deep Semantic Relevance Match in E-commerce Search}
\author {
        Ziyang Liu,\textsuperscript{\rm 1}
        Zhaomeng Cheng, \textsuperscript{\rm 1}
        Yunjiang Jiang \textsuperscript{\rm 1}
        Yue Shang \textsuperscript{\rm 1}
        Wei Xiong \textsuperscript{\rm 1}
        Sulong Xu \textsuperscript{\rm 1}
        Bo Long \textsuperscript{\rm 1}
        Di Jin \textsuperscript{\rm 2} \\
}
\affiliations {
    \textsuperscript{\rm 1} JD.com \\
    \textsuperscript{\rm 2} Tianjin University \\
    liuziyang7@jd.com, chengzhaomeng@jd.com
}
\begin{document}

\maketitle

\begin{abstract}
Result relevance prediction is an essential task of e-commerce search engines to boost the utility of search engines and ensure smooth user experience. The last few years eyewitnessed a flurry of research on the use of Transformer-style models and deep text-match models to improve relevance. However, these two types of models ignored the inherent bipartite network structures that are ubiquitous in e-commerce search logs, making these models ineffective. We propose in this paper a novel \emph{Second-order Relevance}, which is fundamentally different from the previous \emph{First-order Relevance}, to improve result relevance prediction. We design, for the first time, an end-to-end \emph{First-and-Second-order Relevance} prediction model for e-commerce item relevance. The model is augmented by the neighborhood structures of bipartite networks that are built using the information of user behavioral feedback, including clicks and purchases. To ensure that edges accurately encode relevance information, we introduce external knowledge generated from BERT to refine the network of user behaviors. This allows the new model to integrate information from neighboring items and queries, which are highly relevant to the focus query-item pair under consideration. Results of offline experiments showed that the new model significantly improved the prediction accuracy in terms of human relevance judgment. An ablation study showed that the First-and-Second-order model gained a 4.3\% average gain over the First-order model. Results of an online A/B test revealed that the new model derived more commercial benefits compared to the base model.

\end{abstract}

\section{Introduction}
\begin{CJK}{UTF8}{gbsn}
\emph{Semantic Match} is one of the basic tasks for natural language processing and has many real-world applications such as question answering, textual entailment, paraphrase identification, and information retrieval \cite{1,2,3,4}. Unlike simple text matching, semantic similarity match aims to infer the semantic similarity of two sentences rather than the extent of common words between the two. For example, while ``mac pro'' and ``mac lipstick'' look alike, they describe two different items, i.e. computer and lipstick; ``iPad'' and ``apple tablet'' have no common word at all but rather refer to the same item, i.e. tablet computer. While similarity match usually deals with two homogeneous sentences of comparable lengths and expects to match every position of both sentences, semantic relevance match deals with heterogeneous pieces of text such as query and document in ad-hoc information retrieval and expects to match some keywords in documents with queries. As a specific application of ad-hoc information retrieval, e-commerce search serves as a platform to fetch candidate items highly relevant to a given query to satisfy the user's purchase requirement. If a search system returns too many semantically irrelevant items, it will render unpleasant user experience and erode the user's trust and confidence toward the e-commerce platform. Therefore, the semantic relevance estimation is critically important for long-term user satisfaction.

The current research on deep semantic learning can be grouped into two camps: representation-focused and interaction-focused. The representation-focused methods \cite{5,6} typically embed two input sentences separately using two neural networks and computes a relevance score to measure the similarity between the two embeddings. The common similarity measures used include the cosine similarity and negative Jensen-Shannon divergence. On the other hand, the interaction-focused methods \cite{7,8} concatenate two input sentences as input to a single neural network which captures interactions between their features. These two types of methods exploit the semantic relationship between the query and the item text and are effective for static and context-free matching problems. However, the semantic match problem in many real applications is often not context-free, e.g. search logs store plenty of valuable context data\footnote{In this work, context information refers broadly to the neighbor features in a query/item bipartite graph, rather than the more restricted sense of co-occurring items in a search session.} such as the query/item incidence network and the users' historical behavior sequences (view, click, purchase, etc). These types of context information provide the actual semantic content of a query or item, which as ignored by the current semantic match methods.

\begin{figure}[htbp]
\centering
\setlength{\abovecaptionskip}{-0.2cm}
\setlength{\belowcaptionskip}{0cm}
\includegraphics[width=0.5\textwidth]{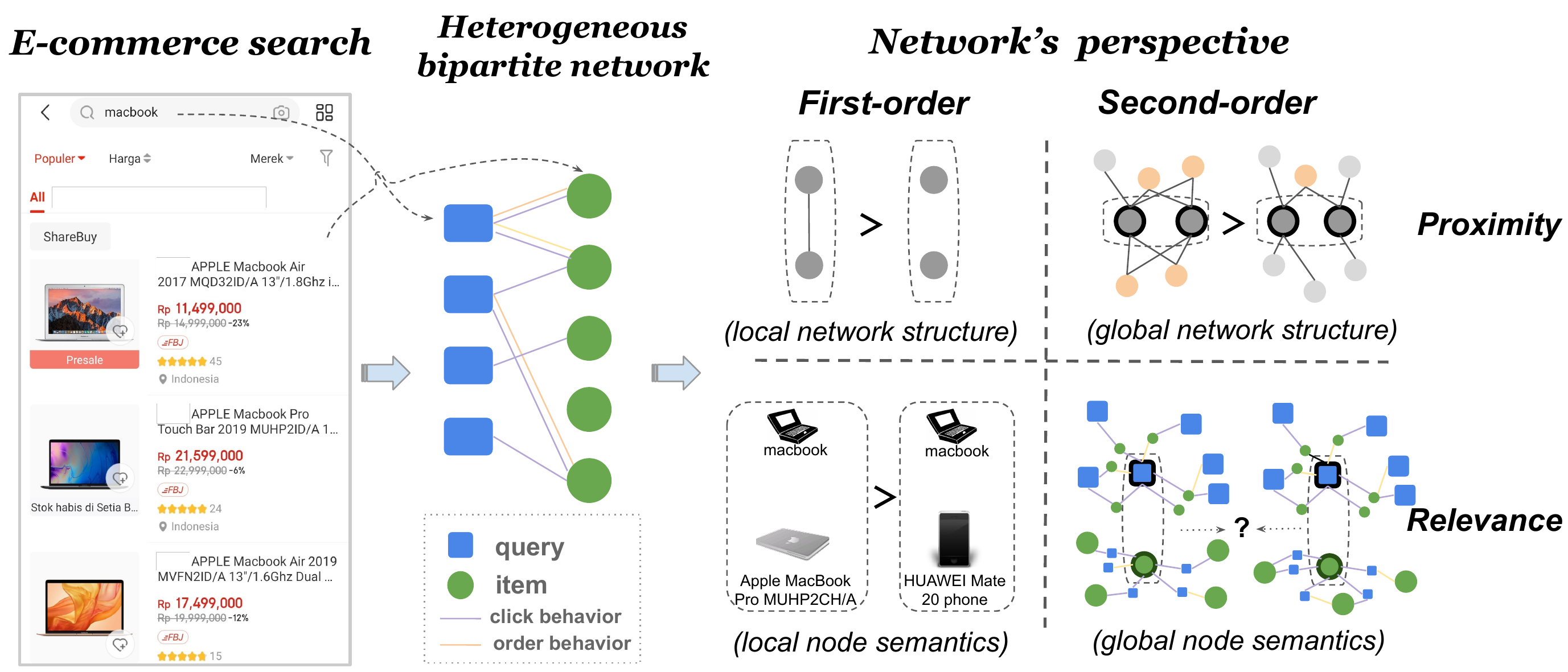}
\hspace{0in}
\caption{In the e-commerce search scene, the search log essentially provides a heterogeneous bipartite network. Similar to 1st/2nd order proximity in network representation learning, we define 1st/2nd order relevance in a semantic match under the e-commerce search scene.}
\label{motivation}
\end{figure}

When exploiting the graphical context information for semantic relevance match, we face the following challenges:

\textbf{C1: Limited supervised information}. 
Although a variety of user behavior signals were recorded in search logs, they are often noisy and misaligned with the search relevance objective: many factors other than relevance may affect a user's final decision such as item popularity, title attractiveness (click-baitiness), and result set diversity. Human labels can provide accurate relevance information, but training an excellent deep model often requires millions or more examples, which are labor-intensive and costly to collect. In short, high quality annotated signals are scarce in our problem domain.

\textbf{C2: Uncertainty in how to integrate context information}. Since user behavioral feedback cannot naively substitute relevance signals, systematic utilization of the massive amounts of search log data has been a central research theme in e-commerce search. Incorporating context information on top of the logged user feedback presents new challenges. Any proposed method should be context-aware and corroborative of the final relevance objective, an under-explored area in the current research.

\textbf{C3: Memory and latency constraints.}
A query phrase and an item's title in an e-commerce search can often be represented in diverse forms. If a text is stored as a unique entity in an online key-value store, it may take 100 Gigabytes of memory (space complexity). Such a large memory usage will result in large models, which in turn will fail to respond quickly to online queries and make the run-time complexity severely limited.

To address these challenges, we propose to study the semantic match problem in dynamically evolving search scenarios. This problem is different from the existing context-free search problems. In particular, we consider that queries and items in the search log constitute a natural heterogeneous bipartite network structure (Fig. \ref{motivation}). In this network, there are two types of nodes (queries and items) and two types of edges (click behaviors and purchase behaviors). Traditional approaches estimate the relevance of two adjacent nodes in this network in isolation. We argue that contextual relevance can be significantly improved by taking into account their neighbors' semantic information.

A central problem in constructing a bipartite network is edge refinement, specifically that 
in the query/item co-occurrence graph. Due to the noisy nature of user behaviors, we cannot rely on them exclusively to build the connection between two nodes. On the other hand, the pre-trained language representation model (e.g. BERT \cite{10}) is equipped with a good semantic understanding capacity. So we choose BERT as the teacher model and extract relevance knowledge which can be used as annotated information (\textbf{C1}). We then use this external knowledge to refine the user behavior network, via adding ignored edges and deleting noisy edges, to ensure a high-quality network structure.

Similar to the concept of first and second order proximity \cite{11,12}, we propose alternative definitions of the first-order relevance and second-order relevance. Previous semantic match research \cite{26,27} only considers the first-order relevance, which is reasonable for context-free semantic match problems, but will lose valuable context information when applied to real-world applications. We argued in \textbf{C2} that both the first and second order relevance should be taken into consideration together for e-commerce search. Therefore, we propose a new model of \textbf{H}eterogeneous \textbf{G}NN for \textbf{S}emantic \textbf{M}atch problem (HG4SM), which can be broadly applied to any search ranking problem that seeks to incorporate the context of real-world applications. The model captures the first-order relevance using a word interaction matrix attached with positional encoding and captures the second-order relevance using the metapath-guided embedding with graph attention scores. To our best knowledge, HG4SM is the first heterogeneous network embedding for the task of search relevance.

Although a query phrase and an item's title may appear in many different forms, the words in these sentences have smaller representation space and are easy to embed. Thus, we use a word distributed representation to depict various queries and titles, which greatly reduces the model's space complexity compared to a document embedding. The above design ensures that our model derives explicit interaction matching signals and reasonable node semantic embeddings so that we only need to employ shallow neural networks to combine all embeddings. Therefore, the whole model has a low time complexity which is suitable to deploy online (\textbf{C3}). We list some related works and compare them with our work in the appendices.
\end{CJK}

\section{Problem Definition}
\newtheorem{myDef}{Definition}
\newtheorem{myTheo}{Theorem}
Here we give: 1) three basic definitions about heterogeneous network and node proximity; 2) and two new definitions about the novel problem of \emph{Second-order Relevance}.
\begin{myDef}
    \textbf{Heterogeneous Network}. Given a network or graph $\mathcal{G} = (\mathcal{V}, \mathcal{E}, \mathcal{A}, \mathcal{R})$ with nodes or vertices set $\mathcal{V} = \left \{v_{1}, v_{2}, \cdots , v_{n}\right \}$ and edges set $\mathcal{E} = \left \{e_{1}, \cdots , e_{m}\right \}$, if the node type's mapping function $\mathcal{V}\rightarrow \mathcal{A}$ and the edge type's mapping function $\mathcal{E}\rightarrow \mathcal{R}$ satisfy the condition: $|\mathcal{A}| + |\mathcal{R}|\geqslant3$, then $\mathcal{G}$ is a heterogeneous network.
\end{myDef}

For the convenience of exploiting the heterogeneous information in a network, we only consider unsigned networks (i.e. no negative edges) with undirected and unweighted edges. In a heterogeneous network, metapath and its corresponding instances are universal concepts and defined as:
\begin{myDef}
    \textbf{Metapath and Metapath Instance}. A metapath $P=a_{1}\overset{r_{1}}{\rightarrow} \cdots \overset{r_{l}}{\rightarrow}a_{l+1}$ represents the path from $a_{1}$ to $a_{l+1}$ successively through $r_{1},\cdots,r_{l}$ ($a_i\in\mathcal{A}$, $r_i\in\mathcal{R}$). A metapath instance $p=\left ( v_{1}, \cdots , v_{l+1}\right )$ is a definite node sequence %founded according to metapath $P$.
    instantiated from metapath $P$.
\end{myDef}

A good network embedding can well preserve the network's structural information, e.g., local network structure and global network structure. The local and global network structures can be respectively represented by the first-order and second-order proximity. Proximity represents two nodes' spatial closeness, where the first-order proximity is the local pairwise proximity and the second-order proximity is the neighbor structure proximity between two nodes.
\begin{myDef}
    \textbf{First-order Proximity and Second-order Proximity}. For two nodes $v_{i}$ and $v_{j}$ in a network, their first-order proximity can be formalized as the function $F_{prx}^{1}(v_{i},v_{j})$ and their second-order proximity can be formalized as the function $F_{prx}^{2}(Nei(v_{i}),Nei(v_{j}))$ where $Nei(.)$ denotes a neighbor set. If there is an edge between $v_{i}$ and $v_{j}$, then the $F_{prx}^{1}(v_{i},v_{j})$'s value is bigger than its value when there is no such edge. If $v_{i}$ and $v_{j}$ share more common neighbors, then the $F_{prx}^{2}(Nei(v_{i}),Nei(v_{j}))$'s value is bigger.
\end{myDef}

For the first-order proximity function, a common design is: 
\begin{equation}
\setlength{\abovedisplayskip}{3pt}
\setlength{\belowdisplayskip}{3pt}
F_{prx}^{1}(v_{i},v_{j})=
\begin{cases}
%1, &\mbox{if there is an edge between $v_{i},v_{j}$;}\\
1, &\mbox{if $e=(v_{i},v_{j})\in\mathcal{E}$;}\\
0, &\mbox{otherwise}
\end{cases}
\end{equation}

For the second-order proximity function, a possible choice can be:
\begin{equation}
\setlength{\abovedisplayskip}{3pt}
\setlength{\belowdisplayskip}{3pt}
    F_{prx}^{2}(Nei(v_{i}),Nei(v_{j}))=\frac{Nei(v_{i})\cap Nei(v_{j})}{Nei(v_{i})\cup Nei(v_{i})}
\end{equation}

Intuitively, even though there is no edge between $v_{1}$ and $v_{2}$, if they share many common neighbors, they should also be very similar to each other. So the second-order proximity is an important supplement to the first-order proximity.

If we view proximity from the perspective of ``path'', we can conclude that: 1) the first-order proximity reveals that the shortest path between $v_{1}$ and $v_{2}$ is a path whose length is 1, and 2) the second-order proximity reveals that the length of their shortest path is 2. Higher-order proximity reveals that the length of their shortest path is greater than 2. With the increase of the path's length, the information intensity will gradually weaken along paths, so higher-order proximity is not considered here. Proximity is also suitable for the heterogeneous network, but the definition of neighbors in the homogeneous network and heterogeneous networks are different. In heterogeneous networks, neighbors are metapath-based and denote those nodes that have the same type as the central node. For example, in a citation network, the author citation relationship can be represented as ``$Author$-$Paper$-$Author$'' and its corresponding metapath is ``$A$-$P$-$A$''. On this metapath, if two authors together cite the same paper, they establish a metapath-based neighbor relationship and tend to have similar research interests. 

In a heterogeneous network, the relevance score represents the semantic closeness of two nodes. Similar to node proximity, we introduce the following new definitions.

\begin{myDef}
    \textbf{First-order Relevance and Second-order Relevance}. When mapping two sentences into two nodes $v_{i}$ and $v_{j}$ in a heterogeneous network $\mathcal{G}$, the first-order relevance and second-order relevance can be separately formalized as the function $F_{rel}^{1}(v_{i},v_{j})$ and  $F_{rel}^{2}(Nei(v_{i}),Nei(v_{j}))$. If $v_{i}$ is semantically more similar to $v_{j}$, then $F_{rel}^{1}(v_{i},v_{j})$ is larger. If $v_{i}$'s neighbors are semantically more similar to $v_{j}$'s neighbors, then $F_{rel}^{2}(Nei(v_{i}),Nei(v_{j}))$ is larger.
\end{myDef}

%Different from the proximity considering the local and global network structures, 
Instead of proximity, our relevance considers local and global node semantics. We argue that both the first-order relevance and second-order relevance are necessary for semantic match. They supplement each other in the same way as the first-order and second-order proximity. A good mechanism should fully consider these two types of relevance and make them cooperate with each other, i.e. the first-order relevance plays a dominant role and the second-order relevance provides important auxiliary information. In other words, when it is difficult to judge whether two nodes are semantically relevant from the textual information of themselves, the second-order relevance which is based on neighbor set can become useful. Finally, we give the definition of the research objective of this work below.

\begin{myDef}
    \textbf{Heterogeneous Network Embedding for Relevance Estimation}. For a heterogeneous network $\mathcal{G}=(\mathcal{V},\mathcal{E},\mathcal{A},\mathcal{R})$, the network embedding for relevance estimation aims to learn each node's low-dimension distributed representation which simultaneously satisfies the need of the first-order relevance and the second-order relevance, and thus preserves both the local and global node semantics. 
\end{myDef}

\section{Model Description}
\begin{figure*}[htbp]
\centering
\includegraphics[width=1.00\textwidth]{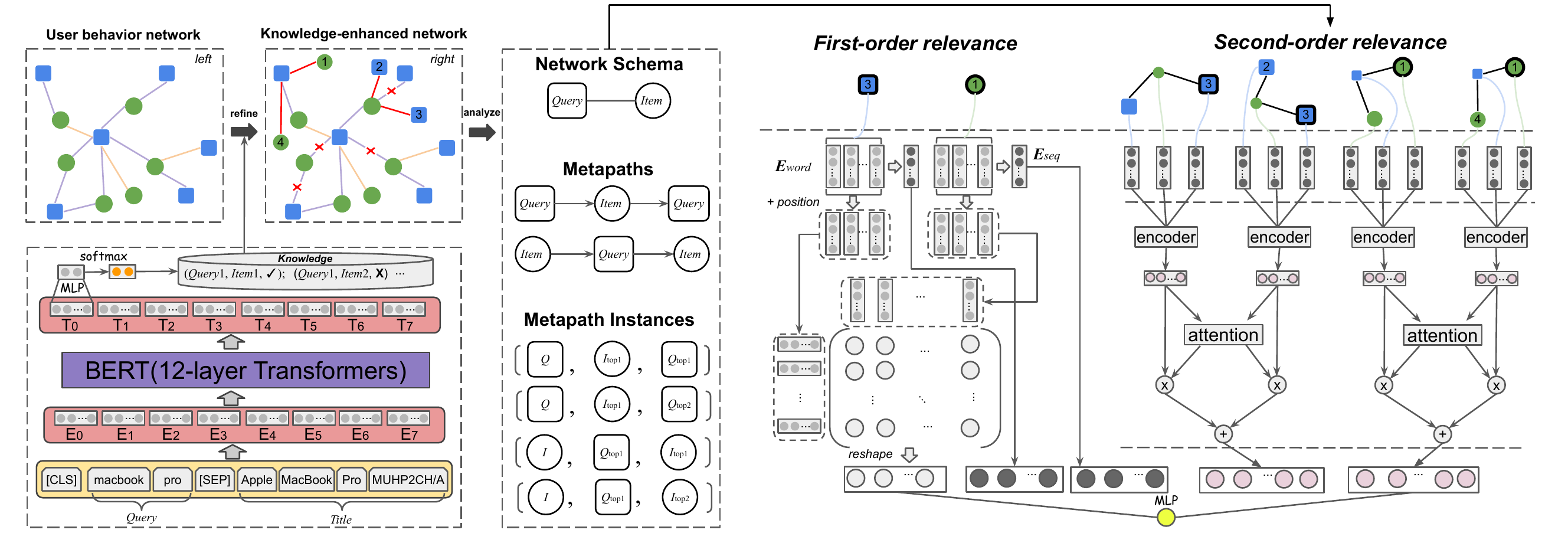}
\hspace{0in}
\setlength{\abovecaptionskip}{-0cm}
\setlength{\belowcaptionskip}{-0.4cm}
\caption{The overall architecture of HG4SM. It mainly includes three steps: knowledge-enhanced network refinement, heterogeneous network analysis and relevance modeling.}
%First, BERT model (red marking) is used to refine click behaviors (purple line), and the edges in the reconstructed network include the original purchase behaviors (yellow line) and the refined click behaviors. In this network, we analyze its schema and employ two types of metapaths (``$Q$-$I$-$Q$'' and ``$I$-$Q$-$I$''). To ensure the quality of neighboring nodes, we select the top-2 relevant nodes as neighbors and constitute four metapath instances based on them. In the model design, the first-order model generates two individual representation-focused embeddings and one interaction-focused embedding; and the second-order model generates two metapath-guided embeddings by applying the node-level encoder and instance-level graph attention. Note that the two sampled nodes (with an identifier of 3 \& 1) and the four sampled metapath instances all come from the knowledge-enhanced user-behavior network.}
\label{model-figure}
\end{figure*}

We propose a complete heterogeneous network embedding architecture for the e-commerce search relevance match problem (Fig. \ref{model-figure}). Concretely, we first need to denoise a user behavior network. We introduce the fine-tuned BERT model of producing external knowledge in order to refine the click behavior edge in the original network. Based on this knowledge-enhanced network derived, we select two most important metapaths. We apply a node-level encoder and a metapath attention unit together to integrate these neighboring nodes' context information into the central node. In addition, considering the e-commerce scene's particularity, we give: 1) a special vocabulary formation rule to preserve the complete semantics of many products or brands, 2) the word-level interaction representation to capture the micro semantics matching signals between the query and title; and 3) the sentence-level semantic representation to capture the macro semantics matching signals between the query and title.

\subsection{Network Construction}
In a real-world search application, queries and items and their relationships naturally form a heterogeneous bipartite network based on the multi-type user behaviors like view, click, and purchase. However, as mentioned earlier, user behaviors are typically biased and noisy. So if one directly conducted embedding learning on the original user behavior network, it would be difficult for the model to estimate explicit relevance relationships. To solve this problem, we introduce external knowledge to refine the user behavior network and then construct a knowledge-enhanced heterogeneous network. Here the external knowledge is provided by the BERT model that is pre-trained on a large text corpus and then fine-tuned on some in-house data. The whole network construction includes the following phrases.

\textbf{Fine-tuning BERT.}
Transformer-based models such as BERT and ERNIE \cite{15} have been preferably used as NLP benchmark in recent years. Here we use the BERT-Base model\footnote{https://github.com/google-research/bert} composed of stacked Transformer units and fine-tune it on some in-house data. The positive and negative examples in the data are human-labeled and cover various categories of items. The fine-tuned BERT is equipped with a high relevance discrimination ability and thus can act as an expert on filtering noisy data.

\textbf{Behavior network formation.}
The user behavior network in the left network of Fig. \ref{model-figure} is built on some user search log (over several months) which records user clicks and purchase behaviors as well as their frequencies. In this network, an edge represents an existing click behavior or purchase behavior between a query-node and an item-node. The edge of purchase behavior is sparse but important. The edge of click behavior is denser but highly noisy, making it difficult to learn a good model and predict reasonable results. Therefore, we introduce BERT to refine this network.

\textbf{Knowledge-enhanced network refinement.}
Given the scarcity of user purchase feedback and noisiness of click feedback, we retain all the original purchase behavior edges while use the external knowledge from BERT to refine the click behaviors. Specifically, we set two different thresholds $\alpha$ and $\beta$, i.e. $\alpha$ is used for judging whether a clicked item is truly relevant to its query, and $\beta$ is used for judging whether an unclicked item is truly irrelevant to its query. This strategy can help remove noises in user behaviors and at the same time extract the missing but crucial relevance signals not captured by user behaviors. To preserve the high-quality neighbor set, for each central node, we rank its neighboring nodes with the priority of ``purchase$\rightarrow$high click$\rightarrow$low click'' and sample %top-$k$ 
top-2 of them as the final neighbor set.

\subsection{Heterogeneous Network Analysis}
In HG4SM, the basic units are query nodes, item nodes and the refined user-behavior-oriented edges between them. Based on this schema, we employ two metapaths, ``$Q$-$I$-$Q$'' and ``$I$-$Q$-$I$'', where ``$Q$'' and ``$I$'' stand for ``$Query$'' and ``$Item$''. These two metapaths correspond closely to the second-order relevance definition defined earlier. Compared to some more complex metapaths (such as ``$Q$-$I$-$Q$-$I$-$Q$'' and ``$I$-$Q$-$I$-$Q$-$I$''), the adopted metapaths in our model is both effective (with less information density loss) and 
computationally efficient. We further choose two instances for each metapath whenever they are available and pad with zero embeddings otherwise
. Take ``$Q$-$I$-$Q$'' as an example, its instances can include ``$Q$-$I_{\mathrm{top1}}$-$Q_\mathrm{{top1}}$'' and ``$Q$-$I_\mathrm{{top1}}$-$Q_\mathrm{{top2}}$'', as shown in Fig.\ref{model-figure}.

\subsection{First-order Relevance Modeling}
The whole framework of the new HG4SM model includes both the first-order relevance modeling and second-order relevance modeling. We first introduce the first-order relevance modeling which captures macro and micro semantic match signals by incorporating both the representation-focused and interaction-focused designs.

\begin{figure}[htbp]
\centering
\vspace{-0.4cm}
\setlength{\abovecaptionskip}{-0.3cm}
\setlength{\belowcaptionskip}{-0.2cm}
\includegraphics[width=0.5\textwidth]{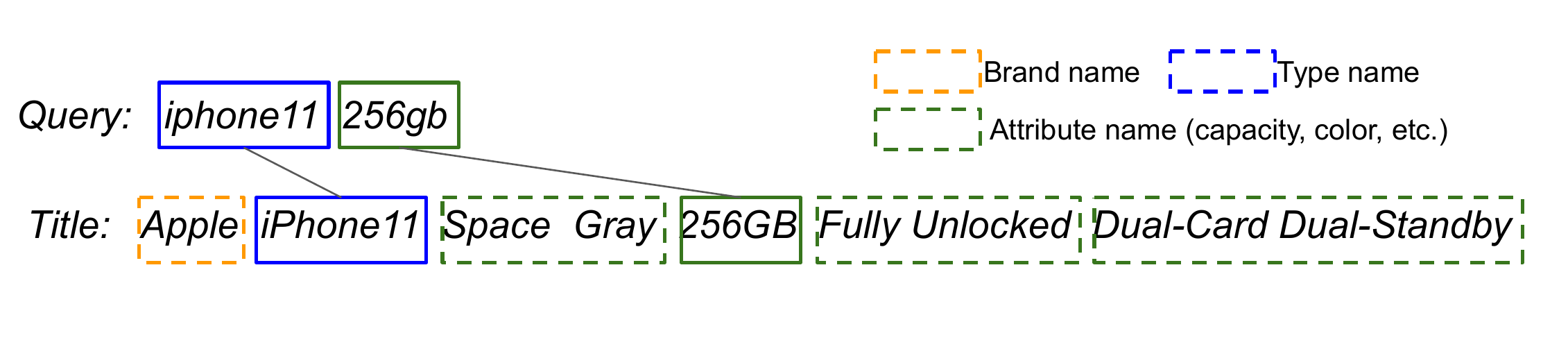}
\hspace{0in}
\caption{An example of the word-level matching signal. The matched and unmatched words are respectively represented by solid and dashed rectangles.}
\label{match-example}
\end{figure}

\textbf{Word embedding in e-commerce scene.} 
In e-commerce applications, it is infeasible to represent queries and items as individual embeddings since their entity space is effectively unbounded. Instead, we adopt word embedding in HG4SM, which dramatically reduces the representation space and deals well with the cold-start situation. Numerous product type names or attribute names (like ``iphone11'' and ``256GB'') exist, but the basic word segmentation based on N-gram or WordPiece word segmentation splits these special names and thus cannot reveal their complete semantics. To adapt to this feature, we first treat single Chinese characters, contiguous numerals or English letters as single words (Fig. \ref{match-example}). We then retain only the high-frequency words in this list. Compared to potentially million queries and billion items, our vocabularies are only in the tens of thousands, which saves memory consumption and lookup time of id and embedding by a large margin. We represent each word with a $d$-dimensional vector and train these word vectors or embeddings using Word2Vec \cite{24}. We use $\bm{E}_{Q}^{i}$ (and $\bm{E}_{I}^{i}$) to denote the $i$-th word's embedding of query $Q$ (and the item's title $I$).

\textbf{Macro matching element.} Most of the first-order relevance methods are either representation-focused (capturing macro matching signals) or interaction-focused (capturing micro matching signals). We employ their mixture in the HG4SM's first-order relevance modeling. For the representation-focused part, take query $Q$ with $l_Q$ words as an example, its sequence embedding $\bm{E}_{seq}^{Q}$ is obtained by calculating the element-wise mean value of $\left \{\bm{E}_{Q}^{i}\right \}$:
\begin{equation}
\setlength{\abovedisplayskip}{3pt}
\setlength{\belowdisplayskip}{3pt}
\bm{E}_{seq}^{Q}=\frac{1}{l_Q}\sum_{i=1}^{l_Q}\bm{E}_{Q}^{i}
\end{equation}
The above $\bm{E}_{seq}^{Q}$ depicts the whole semantic information of query $Q$, so that it can be viewed as a simplified version of representation-focused embedding.

%\textbf{Relationship with other representation-focused method.} Deep Structured Semantic Models (DSSM)\cite{dssm} is a classical work for web search task. It transforms two documents into corresponding semantic embedding vectors by word hashing and deep neural networks, then calculates their relevance score by cosine similarity. The inherent representation ideas in DSSM and HG4SM are the same, i.e., separate sentence representation and similarity calculation. While, their difference is: 1. for separate sentence representation, HG4SM directly uses Eq. (3) (not neural networks) to represent sentence's semantics so that it can reduce time complexity; 2. for similarity calculation, HG4SM uses neural networks (not cosine similarity) to measure query and item's relevance in the stage of embedding fusion (we will introduce it in subsection \emph{D}) so that it can integrate both of first-order embedding and second-order embedding into relevance estimation.

%\textbf{Interaction-focused part.}
\textbf{Micro matching element.} To capture micro matching signals, we need to model the word-level interaction information \cite{7} between queries and titles. Suppose the sequence lengths of query $Q$ and title $I$ are $l_Q$ and $l_I$, respectively, we build an interaction matrix $\bm{M}_{int}$:
\begin{equation}
\bm{M}_{int} = [\bm{E}_{Q}^{1}, ..., \bm{E}_{Q}^{l_Q}]^{\mathrm{T}} \times [\bm{E}_{I}^{1}, ..., \bm{E}_{I}^{l_I}]
\end{equation}

The interaction embedding is derived by reshaping $\bm{M}_{int}$ into a one-dimension vector:
\begin{equation}
\setlength{\abovedisplayskip}{3pt}
\setlength{\belowdisplayskip}{3pt}
\bm{E}_{int}=reshape(\bm{M}_{int})
\end{equation}

\textbf{Position encoding.} Besides, considering the sequential structure of texts, we further add a position embedding to each word embedding before calculating the correlation matrix. The position embedding is set as a trainable embedding vector and has the same dimension as the word embedding.

\subsection{Second-order Relevance Modeling}
Most semantic match methods focus on the first-order relevance, but ignore the second-order relevance (which integrates the neighbor information on metapaths into the central node and is essentially important in many real context-aware scenes). A complete semantic relevance estimation model should integrate them together. Here we consider how to generate second-order relevance embeddings so that it can incorporate context information in the network to enrich and improve the central nodes' semantics. In general, the second-order relevance model consists of a node-level encoder and a metapath instance-level aggregator.

\textbf{Node-level encoder.}
Metapath instance bridges the communication gap between heterogeneous nodes and thereby can infer the node's global semantic embedding (rather than the local semantic embedding). For each metapath instance, to derive the global node semantics, we integrate the neighboring node embedding into the central node embedding with mean encoder. Take the instance ``$Q$-$I_\mathrm{{top1}}$-$Q_\mathrm{{top1}}$'' as an example, its corresponding embedding is:
\begin{equation}
\setlength{\abovedisplayskip}{3pt}
\setlength{\belowdisplayskip}{3pt}
\bm{E}_{Q-I_\mathrm{{top1}}-Q_\mathrm{{top1}}}=\mathrm{MEAN}\left \{\bm{E}_{seq}^{Q}, \bm{E}_{seq}^{I_{\mathrm{top1}}},\bm{E}_{seq}^{Q_{\mathrm{top1}}}\right \}
\end{equation}

\textbf{Instance-level aggregator with graph attention.}
Different meatapath instances convey different information, so they should have different effects on the final metapath's embedding. However, the mapping relationship between the metapath instance's embedding and the metapath's embedding is unknown. To learn their relationship automatically, we introduce a ``graph attention'' mechanism to the formation progress of the metapath's embedding. The attention mechanism enables the model to learn a weight distribution and assign different weights to different components, which has been successfully applied in computer vision and natural language processing \cite{23}. Here we introduce graph attention to represent the mapping relationship between metapath and its instances. The final metapath's embedding is then obtained by accumulating the embeddings of each metapath's instances with attention scores. Take metapath ``$Q$-$I$-$Q$'' as an example, its corresponding embedding is defined as:
\begin{align} \nonumber
\setlength{\abovedisplayskip}{3pt}
\setlength{\belowdisplayskip}{3pt}
&\bm{E}_{Q-I-Q}= \\
&\sigma (\mathrm{Att}_{1}\cdot \bm{E}_{Q-I_\mathrm{{top1}}-Q_\mathrm{{top1}}}+\mathrm{Att}_{2}\cdot \bm{E}_{Q-I_\mathrm{{top1}}-Q_\mathrm{{top2}}})
\end{align}
where $\sigma(.)$ is the activation function. Though $\mathrm{Att}_{i}$ can be set as a fixed value, we adopt a more flexible way, i.e., we use a single-layer neural network to learn it automatically. Specifically, we feed the concatenation of embeddings of the central node and the metapath instances into a one-layer neural network and output an attention distribution in the softmax layer:
\begin{equation}
\setlength{\abovedisplayskip}{3pt}
\setlength{\belowdisplayskip}{3pt}
\bm{E}_\mathrm{concat}=\mathrm{CONCAT}(\bm{E}_{seq}^{Q}, \bm{E}_{Q-I_\mathrm{{top1}}-Q_\mathrm{{top1}}})
\end{equation}
\begin{equation}
\setlength{\abovedisplayskip}{3pt}
\setlength{\belowdisplayskip}{3pt}
\mathrm{Att}_{i}=\mathrm{softmax}(\sigma (W_{att}*\bm{E}_\mathrm{concat}+b))
\end{equation}
where $W_{att}$ is a 1*2\emph{d} trainable vector, shared across all metapaths.

\textbf{Embedding fusion.}
Based on the first-order and second-order relevance modeling, three types of embeddings can be generated, including the representation-
focused,	interaction-focused	and metapath-guided embeddings. To combine them, we concatenate these three embeddings together, feed it to the deep neural networks, and output a relevance score. To alleviate the model's high time-cost, we use efficient three-layer DNNs without considering more complex neural network structures such as CNN and LSTM.

\section{Experiments}
We now present experimental results. We first introduce the baseline methods, performance metrics, final comparison results, ablation study, and online performance. We discuss the datasets used, implementation details, and parameters sensitivity experiment in the appendices.

\subsection{The Baselines and Metrics}
The HG4SM model is compared to several existing state-of-the-art semantic models. To facilitate their implementation, we use the open-source codebase packages which are as follows. \\
\textsuperscript{\dag}: Nine methods are implemented using MatchZoo \cite{25}, which is a toolkit for deep text matching, based on TensorFlow and Keras. We compare these typical methods \cite{28,29,30,31,32,33} in MatchZoo with HG4SM. We use the default hyper-parameter setting for all the methods in MatchZoo\footnote{https://github.com/NTMC-Community/MatchZoo}.\\
\textsuperscript{\S}: Two other baselines DSSM \cite{5} and ESIM \cite{8} are also included in the comparison. DSSM is a classical representation-focused method using pairwise examples to train the model. ESIM is a successful interaction-focused method which incorporates the syntactic parsing information into chain LSTM for natural language inference.

To comprehensively measure the performance of our HG4SM and these baseline methods, we use six metrics, including: 1) Area Under the receiver operating characteristic Curve, 2) Accuracy, 3) Precision, 4) Recall, 5) F1-score and 6) False Negative Rate (since it is more serious than False Positive Rate for e-commerce ranking). They are respectively denoted by AUC, Acc, Pre, Recall, F1 and FNR for short. For the first five metrics, the higher the metric value, the better the model's performance. For FNR, the lower its value, the better the model's performance. Note that AUC often serves as the most important metric while the others also provide auxiliary supports for our analysis.

\begin{table*}[ht]\scriptsize
\setlength{\abovecaptionskip}{-0cm} 
\setlength{\belowcaptionskip}{-0cm}
\caption{Comparison on the in-house data. The best results are bolded. `Rep' denotes the representation-focused method, `Int' the interaction-focused method, and `Both' their mixture. `Triple' denotes `Both' with `HIN'. `$\blacktriangleleft$' denotes the most important metric in the real application. `(-)' represents that the lower value corresponds to better performance. Here the used all-categories data is the sampled 10 million data.}
\begin{center}
\begin{tabular}{cccccccccccccc}
\toprule[0.8pt]
\multirow{2}*{Types} &\multirow{2}*{Models}   & \multicolumn{6}{c}{Mobile-phone} & \multicolumn{6}{c}{All-categories (sampled)}   \\
 & & AUC$\blacktriangleleft$ & Acc & Prec & Recall & F1 & FNR(-) & AUC$\blacktriangleleft$ & Acc & Prec & Recall & F1 & FNR(-) \\ \midrule[0.8pt]
Rep &DSSM\textsuperscript{\S} &0.6246 &0.5304 &0.5300 &\textbf{0.9977} &0.6923 &0.9953 &0.8219 &0.7686 &0.7686 &\textbf{1.0000} &0.8691 &1.0000 \\
Rep &MVLSTM\textsuperscript{\dag} &0.8602 &0.7752 &0.7433 &0.8791 &0.8055 &0.3416 &0.7877 &0.8052 &0.8090 &0.9786 &0.8857 &0.7802	\\
Rep &ARC-I\textsuperscript{\dag}	&0.8343 &0.7580 &0.7210 &0.8858 &0.7949 &0.3857 &0.6919	&0.7809	&0.7814	&0.9941	&0.8750 &0.9388 \\
Int &DRMM\textsuperscript{\dag}  &0.6720 &0.5771 &0.5642 &0.8850 &0.6891 &0.7692 &0.6781 &0.7765 &0.7803 &0.9888 &0.8722 &0.9401 \\
Int &MatchPyramid\textsuperscript{\dag}	&0.7826 &0.6806 &0.6422 &0.8957 &0.7481 &0.5615 &0.7859	&0.7911	&0.7961	&0.9802	&0.8786 &0.8475\\
Int &ARC-II\textsuperscript{\dag}	&0.8128 &0.7411 &0.6982 &0.9001 &0.7864 &0.4377 &0.7606  &0.7878  &0.7871  &0.9938 &0.8784 &0.9076\\
Int &K-NRM\textsuperscript{\dag}	&0.7462 &0.6438 &0.6102 &0.9057 &0.7291 &0.6510 &0.7314 &0.7799 &0.7853 &0.9837 &0.8733 &0.9081 \\
Int &DRMM-TKS\textsuperscript{\dag}	&0.7678 &0.6738 &0.6417 &0.8692 &0.7383 &0.5462 &0.7793 &0.7943 &0.8053 &0.9672 &0.8789 &0.7893 \\	
Int &Conv-KNRM\textsuperscript{\dag} &0.8369 &0.7575 &0.7339 &0.8504 &0.7879 &0.3469 &0.8029 &0.7894 &0.7896 &0.991 &0.8789 &0.8913 \\
Int &ESIM\textsuperscript{\S} &0.8056 &0.7520 &0.7526 &0.8029 &0.7769 &0.3073 &0.7987 &0.7580 &0.7580 &1.0000 &0.8623 &1.0000 \\
Both &Duet\textsuperscript{\dag}	&0.7693 &0.6023 &0.5731 &0.9752 &0.7219 &0.8173 &0.7968 &0.7812 &0.7806 &0.9965 &0.8754 &0.9458 \\
Triple &HG4SM &\textbf{0.8785} &\textbf{0.8013} &\textbf{0.7711} &0.8883 &\textbf{0.8256} &\textbf{0.2966} &\textbf{0.8758} &\textbf{0.8559} &\textbf{0.8956} &0.9206 &\textbf{0.9079} &\textbf{0.3625} \\
\bottomrule[0.8pt]
\end{tabular}
\end{center}
\label{table-compare}
\end{table*}

\begin{table*}[ht]\scriptsize
\setlength{\abovecaptionskip}{-0cm} 
\setlength{\belowcaptionskip}{-0cm}
\caption{Ablation study. Best results are bolded. `Rep', `Int' and `HIN' are single-component models, and `Rep+Int', `Int+HIN' and `Rep+HIN' both-component models. Bolded numbers are the best result under the current metric. Note: here all-categories is complete 170 million data
different from the 10 million data in the Table \ref{table-compare}.}
\begin{center}
\begin{tabular}{cclccccclccccc}
\toprule[0.8pt]
\multirow{2}*{Orders} &\multirow{2}*{Models}   & \multicolumn{6}{c}{Mobile-Phone} & \multicolumn{6}{c}{All-categories (complete)}   \\
& & AUC$\blacktriangleleft$ & Acc & Prec & Recall & F1 & FNR(-) & AUC$\blacktriangleleft$ & Acc & Prec & Recall & F1 & FNR(-) \\ \midrule[0.8pt]
1st &Rep	&0.8537	&0.7719	&0.7368 &\textbf{0.8856}	&0.8044	&0.3560 &0.8067 &0.7978 &0.8855 &0.8474 &0.8660 &0.3697 \\
1st &Int &0.8595 &0.7727 &0.7405 &0.8787 &0.8037 &0.3464	&0.8289 &0.8534 &0.8770 &0.9421 &0.9084 &0.4459 \\
2nd &HIN &0.8430	&0.7929	&0.7667	&0.8751	&0.8173	&0.2995	&0.8500	&0.8592	&0.8852	&0.9392	&\textbf{0.9114}	&0.4110 \\
1st &Rep+Int &0.8638 &0.7795	&0.7481	&0.8797	&0.8086	&0.3331	&0.8776	&0.8503	&0.8707	&\textbf{0.9465}	&0.9070	&0.4743 \\
1st \& 2nd &Int+HIN &0.8761 &0.8009 &\textbf{0.7799} &0.8691 &0.8221 &\textbf{0.2758}	&0.8824	&0.8585	&0.8941	&0.9263	&0.9099	&0.3705 \\
1st \& 2nd &Rep+HIN &0.8656	&0.7956	&0.7708	&0.8736	&0.8190	&0.2922	&0.8750	&0.8576	&0.8928	&0.9266	&0.9094	&0.3753 \\
1st \& 2nd &HG4SM &\textbf{0.8786} &\textbf{0.8025} &0.7790	&0.8754 &\textbf{0.8244} &0.2794 &\textbf{0.8862} &\textbf{0.8597} &\textbf{0.8949}	&0.9270	&0.9107 &\textbf{0.3673}\\
\bottomrule[0.8pt]
\end{tabular}
\end{center}
\label{table-ablation}
\end{table*}

\subsection{Comparisons}
We compare HG4SM with eleven state-of-the-art deep semantic matching methods using the in-house e-commerce search log data. The results are shown in Table \ref{table-compare}. Because some baseline methods (e.g., DRMM and ESIM) have relatively high time complexities, we sample ten million training data from the all-categories dataset. For fairness, all methods including our HG4SM are trained on these data.

As shown in Table \ref{table-compare}, HG4SM nearly always outperforms the other methods compared, across all six metrics. More specifically,
\begin{itemize}
\item Compared to the second best method, HG4SM obtains 1.8\% (and 5.4\%) gains under AUC in the mobile-phone- dataset (and all-categories dataset). Furthermore, HG4SM achieves the best (smallest) FNR on both these datasets. This implies that HG4SM has a high discrimination power on negative examples, so that it can return a list of satisfactory items which are relevant to the user's shopping needs.
\item The collected training data have imbalanced classes (i.e. the positive examples are far more than the negative examples), making model learning a challenge. Most of the methods compared, such as DSSM, are vulnerable to the class imbalance and often fail to correctly estimate many testing examples in this case. Fortunately, our HG4SM learns explicit node semantics benefiting from the neighboring node's effect, so that it has a robust performance even though the training data are highly imbalanced.
\end{itemize}

\subsection{Ablation Study}
To further examine the importance of each component in the HG4SM model, we remove one or two components from HG4SM at a time and examine how the components affect the overall performance.% of the method. 
We have the following empirical observation and analysis of the results, shown in Table \ref{table-ablation}:
\begin{itemize}
\item In general, for the three submodels of HG4SM (including the representation-focused	component, the interaction-focused component, and both), the both-component setting often outperforms each single-component setting but worse than the triple-component setting by introducing HIN (i.e. HG4SM). It demonstrates that introducing HIN helps the model learn comprehensive knowledge and thus gives an explicit estimation.
\item The introduction of second-order relevance modeling can provide stable improvement (e.g. 1.0\%$\sim$1.6\% of AUC gains) to every first-order solution. This demonstrates that applying HIN to the semantic match can effectively exploit the neighboring nodes' information contained in user-behavior networks, benefiting the final relevance estimation between central nodes.
\end{itemize}

\subsection{Online Performance}
To further evaluate HG4SM's performance in the real search scene, we deploy it to an online e-commerce search platform and report its A/B test results in Table \ref{online}. Four widely-used online business metrics are adopted: 1) Conversion Rate (\textbf{CVR}): Total order number / total click number; 2) User Conversion Rate (\textbf{UCVR}): Total order number / search user number; 3) \textbf{UV-value}: Total Gross Merchandise Volume / total user number; and 4) Revenue Per Mile (\textbf{RPM}): 1000 * search GMV / search number.

\begin{table}[ht]\footnotesize
\caption{HG4SM's online performance under price sort mode and default sort mode. All reported results are observed in at least ten consecutive days. The online baseline method is a feature-based GBDT model trained on a human-labeled dataset containing 73 hand-crafting features.}
\begin{center}
\begin{tabular}{ccccc}
\toprule[0.8pt]
\multirow{2}*{Metrics} &\multicolumn{2}{c}{Price Sort} &\multicolumn{2}{c}{Default Sort}\\
 &Improvement &P-value &Improvement &P-value\\
\midrule[0.8pt]
UV-value &5.713\% &3.20e-2 &0.5013\%&1.10e-1\\
UCVR &1.540\% &7.81e-2 &0.3058\%&1.75e-2\\
CVR &1.829\% &1.01e-2 &0.1218\%&1.60e-1\\
RPM &5.587\% &3.03e-2 &0.6886\%&2.32e-2\\
\bottomrule[0.8pt]
\end{tabular}
\end{center}
\label{online}
\end{table}

The results of A/B tests show that our HG4SM outperforms the existing online DNN model in this platform on all of the business metrics. For example, HG4SM improves 5.7\% and 0.5\% of UV-values under both price sort and default sort. It indicates that 1) our HG4SM model has a low time complexity, which makes it easy to cooperate with other online serial models, and 2) HG4SM provides accurate relevance estimation between queries and items, so that it provides users efficient and intelligent search experiences.

\section{Conclusion}
In this paper, we studied the semantic relevance match problem in e-commerce search. In reference to the previous semantic match research using first-order relevance modeling, we proposed a novel idea to combine first-order and second-order relevance match. Based on this new idea, we employed a heterogeneous network embedding to exploit the potential context information in the ``query-item'' heterogeneous bipartite network. Compared to the current state-of-the-art methods, our novel HG4SM model showed a robust prediction performance. The ablation study verified that the addition of the second-order relevance modeling can significantly improve the performance of the method using the traditional first-order relevance alone. Finally, we applied HG4SM to our in-house e-commerce platform by deploying it to the online search system, which significantly improved the user's search experience.

\bibliography{aaai.bib}

\begin{thebibliography}{23}
\providecommand{\natexlab}[1]{#1}
\providecommand{\url}[1]{\texttt{#1}}
\providecommand{\urlprefix}{URL }
\expandafter\ifx\csname urlstyle\endcsname\relax
  \providecommand{\doi}[1]{doi:\discretionary{}{}{}#1}\else
  \providecommand{\doi}{doi:\discretionary{}{}{}\begingroup
  \urlstyle{rm}\Url}\fi

\bibitem[{Berger et~al.(2000)Berger, Caruana, Cohn, Freitag, and Mittal}]{1}
Berger, A.; Caruana, R.; Cohn, D.; Freitag, D.; and Mittal, V. 2000.
\newblock Bridging the lexical chasm: statistical approaches to answer-finding.
\newblock In \emph{Proceedings of the 23rd annual international ACM SIGIR
  conference on Research and development in information retrieval}, 192--199.

\bibitem[{Chen et~al.(2017)Chen, Zhu, Ling, Wei, Jiang, and Inkpen}]{8}
Chen, Q.; Zhu, X.; Ling, Z.-H.; Wei, S.; Jiang, H.; and Inkpen, D. 2017.
\newblock Enhanced LSTM for Natural Language Inference.
\newblock In \emph{Proceedings of the 55th Annual Meeting of the Association
  for Computational Linguistics (Volume 1: Long Papers)}, 1657--1668.

\bibitem[{Dagan, Glickman, and Magnini(2005)}]{2}
Dagan, I.; Glickman, O.; and Magnini, B. 2005.
\newblock The PASCAL recognising textual entailment challenge.
\newblock In \emph{Machine Learning Challenges Workshop}, 177--190. Springer.

\bibitem[{Devlin et~al.(2019)Devlin, Chang, Lee, and Toutanova}]{10}
Devlin, J.; Chang, M.-W.; Lee, K.; and Toutanova, K. 2019.
\newblock BERT: Pre-training of Deep Bidirectional Transformers for Language
  Understanding.
\newblock In \emph{Proceedings of the 2019 Conference of the North American
  Chapter of the Association for Computational Linguistics: Human Language
  Technologies, Volume 1 (Long and Short Papers)}, 4171--4186.

\bibitem[{Dolan et~al.(2004)Dolan, Quirk, Brockett, and Dolan}]{3}
Dolan, W.; Quirk, C.; Brockett, C.; and Dolan, B. 2004.
\newblock Unsupervised construction of large paraphrase corpora: Exploiting
  massively parallel news sources.
\newblock In \emph{Proceedings of the 20th International Conference on
  Computational Linguistics}.

\bibitem[{Fan et~al.(2017)Fan, Pang, Hou, Guo, Lan, and Cheng}]{25}
Fan, Y.; Pang, L.; Hou, J.; Guo, J.; Lan, Y.; and Cheng, X. 2017.
\newblock Matchzoo: A toolkit for deep text matching.
\newblock In \emph{Neu-IR: The SIGIR 2017 Workshop on Neural Information
  Retrieval}.

\bibitem[{Guo et~al.(2016)Guo, Fan, Ai, and Croft}]{28}
Guo, J.; Fan, Y.; Ai, Q.; and Croft, W.~B. 2016.
\newblock A deep relevance matching model for ad-hoc retrieval.
\newblock In \emph{Proceedings of the 25th ACM International on Conference on
  Information and Knowledge Management}, 55--64.

\bibitem[{Hu et~al.(2014)Hu, Lu, Li, and Chen}]{30}
Hu, B.; Lu, Z.; Li, H.; and Chen, Q. 2014.
\newblock Convolutional Neural Network Architectures for Matching Natural
  Language Sentences.
\newblock In \emph{Advances in Neural Information Processing Systems 27: Annual
  Conference on Neural Information Processing Systems 2014, December 8-13 2014,
  Montreal, Quebec, Canada}, 2042--2050.

\bibitem[{Huang et~al.(2013)Huang, He, Gao, Deng, Acero, and Heck}]{5}
Huang, P.-S.; He, X.; Gao, J.; Deng, L.; Acero, A.; and Heck, L. 2013.
\newblock Learning deep structured semantic models for web search using
  clickthrough data.
\newblock In \emph{Proceedings of the 22nd ACM international conference on
  Information \& Knowledge Management}, 2333--2338.

\bibitem[{Jiang et~al.(2019)Jiang, Shang, Li, Yang, Tang, Ma, Xiao, and
  Zhao}]{27}
Jiang, Y.; Shang, Y.; Li, R.; Yang, W.-Y.; Tang, G.; Ma, C.; Xiao, Y.; and
  Zhao, E. 2019.
\newblock A unified neural network approach to e-commerce relevance learning.
\newblock In \emph{Proceedings of the 1st International Workshop on Deep
  Learning Practice for High-Dimensional Sparse Data}, 10. ACM.

\bibitem[{Li(2011)}]{4}
Li, H. 2011.
\newblock Learning to rank for information retrieval and natural language
  processing.
\newblock \emph{Synthesis Lectures on Human Language Technologies} 4(1):
  1--113.

\bibitem[{Mikolov et~al.(2013)Mikolov, Chen, Corrado, and Dean}]{24}
Mikolov, T.; Chen, K.; Corrado, G.; and Dean, J. 2013.
\newblock Efficient Estimation of Word Representations in Vector Space.
\newblock In \emph{1st International Conference on Learning Representations,
  {ICLR} 2013, Scottsdale, Arizona, USA, May 2-4, 2013, Workshop Track
  Proceedings}.

\bibitem[{Mitra, Diaz, and Craswell(2017)}]{32}
Mitra, B.; Diaz, F.; and Craswell, N. 2017.
\newblock Learning to Match using Local and Distributed Representations of Text
  for Web Search.
\newblock In \emph{Proceedings of the 26th International Conference on World
  Wide Web, {WWW} 2017, Perth, Australia, April 3-7, 2017}, 1291--1299. {ACM}.

\bibitem[{Nigam et~al.(2019)Nigam, Song, Mohan, Lakshman, Ding, Shingavi, Teo,
  Gu, and Yin}]{26}
Nigam, P.; Song, Y.; Mohan, V.; Lakshman, V.; Ding, W.; Shingavi, A.; Teo,
  C.~H.; Gu, H.; and Yin, B. 2019.
\newblock Semantic product search.
\newblock In \emph{Proceedings of the 25th ACM SIGKDD International Conference
  on Knowledge Discovery \& Data Mining}, 2876--2885.

\bibitem[{Pang et~al.(2016{\natexlab{a}})Pang, Lan, Guo, Xu, Wan, and
  Cheng}]{7}
Pang, L.; Lan, Y.; Guo, J.; Xu, J.; Wan, S.; and Cheng, X. 2016{\natexlab{a}}.
\newblock Text matching as image recognition.
\newblock In \emph{Proceedings of the 30th AAAI Conference on Artificial
  Intelligence}.

\bibitem[{Pang et~al.(2016{\natexlab{b}})Pang, Lan, Guo, Xu, Wan, and
  Cheng}]{29}
Pang, L.; Lan, Y.; Guo, J.; Xu, J.; Wan, S.; and Cheng, X. 2016{\natexlab{b}}.
\newblock Text Matching as Image Recognition.
\newblock In \emph{Proceedings of the Thirtieth {AAAI} Conference on Artificial
  Intelligence, February 12-17, 2016, Phoenix, Arizona, {USA}}, 2793--2799.

\bibitem[{Sun et~al.(2019)Sun, Wang, Li, Feng, Chen, Zhang, Tian, Zhu, Tian,
  and Wu}]{15}
Sun, Y.; Wang, S.; Li, Y.; Feng, S.; Chen, X.; Zhang, H.; Tian, X.; Zhu, D.;
  Tian, H.; and Wu, H. 2019.
\newblock ERNIE: Enhanced representation through knowledge integration.
\newblock \emph{arXiv preprint arXiv:1904.09223} .

\bibitem[{Tang et~al.(2015)Tang, Qu, Wang, Zhang, Yan, and Mei}]{11}
Tang, J.; Qu, M.; Wang, M.; Zhang, M.; Yan, J.; and Mei, Q. 2015.
\newblock Line: Large-scale information network embedding.
\newblock In \emph{Proceedings of the 24th international conference on world
  wide web}, 1067--1077.

\bibitem[{Vaswani et~al.(2017)Vaswani, Shazeer, Parmar, Uszkoreit, Jones,
  Gomez, Kaiser, and Polosukhin}]{23}
Vaswani, A.; Shazeer, N.; Parmar, N.; Uszkoreit, J.; Jones, L.; Gomez, A.~N.;
  Kaiser, {\L}.; and Polosukhin, I. 2017.
\newblock Attention is all you need.
\newblock In \emph{Advances in neural information processing systems},
  5998--6008.

\bibitem[{Wan et~al.(2016{\natexlab{a}})Wan, Lan, Guo, Xu, Pang, and Cheng}]{6}
Wan, S.; Lan, Y.; Guo, J.; Xu, J.; Pang, L.; and Cheng, X. 2016{\natexlab{a}}.
\newblock A deep architecture for semantic matching with multiple positional
  sentence representations.
\newblock In \emph{Proceedings of the 30th AAAI Conference on Artificial
  Intelligence}.

\bibitem[{Wan et~al.(2016{\natexlab{b}})Wan, Lan, Guo, Xu, Pang, and
  Cheng}]{31}
Wan, S.; Lan, Y.; Guo, J.; Xu, J.; Pang, L.; and Cheng, X. 2016{\natexlab{b}}.
\newblock A Deep Architecture for Semantic Matching with Multiple Positional
  Sentence Representations.
\newblock In \emph{Proceedings of the Thirtieth {AAAI} Conference on Artificial
  Intelligence, February 12-17, 2016, Phoenix, Arizona, {USA}}, 2835--2841.

\bibitem[{Wang, Cui, and Zhu(2016)}]{12}
Wang, D.; Cui, P.; and Zhu, W. 2016.
\newblock Structural deep network embedding.
\newblock In \emph{Proceedings of the 22nd ACM SIGKDD international conference
  on Knowledge discovery and data mining}, 1225--1234.

\bibitem[{Xiong et~al.(2017)Xiong, Dai, Callan, Liu, and Power}]{33}
Xiong, C.; Dai, Z.; Callan, J.; Liu, Z.; and Power, R. 2017.
\newblock End-to-End Neural Ad-hoc Ranking with Kernel Pooling.
\newblock In \emph{Proceedings of the 40th International {ACM} {SIGIR}
  Conference on Research and Development in Information Retrieval, Shinjuku,
  Tokyo, Japan, August 7-11, 2017}, 55--64. {ACM}.

\end{thebibliography}
\end{document}